\documentclass[preprints,article,accept,moreauthors,fleqn]{Definitions/mdpi} 
\usepackage{amssymb}
\usepackage{amsmath}
\usepackage{array}
\firstpage{1} 
\pubvolume{xx}
\issuenum{1}
\articlenumber{5}
\pubyear{2019}
\copyrightyear{2019}
\history{Received: date; Accepted: date; Published: date}

\Title{Atomic data for calculation of the intensities of Stark components of excited hydrogen atoms in fusion plasmas}

\Author{Oleksandr Marchuk$^1$\orcidA{}, David R. Schultz$^2$ and Yuri Ralchenko$^3$\orcidC{}}

\AuthorNames{Oleksandr Marchuk, David R. Schultz, Yuri Ralchenko}

\address{%
$^{1}$ \quad Forschungszentrum J{\"u}lich GmbH, Institut f{\"u}r Energie- und Klimaforschung - Plasmaphysik, Partner of the Trilateral Euregio Cluster (TEC), 52425 J{\"u}lich, Germany\\
$^{2}$ \quad Department of Astronomy and Planetary Science, Northern Arizona University, Flagstaff, AZ 86011,  United States of America\\
$^{3}$ \quad Atomic Spectroscopy Group, National Institute of Standards and Technology, Gaithersburg, MD 20899-8422, United States of America}

\corres{Corresponding author. Email: o.marchuk@fz-juelich.de.}

\abstract{ Motional Stark effect (MSE) spectroscopy represents a unique diagnostic tool capable of determining the magnitude of the magnetic field and its direction in the core of fusion plasmas. The primary excitation channel for fast hydrogen atoms in injected neutral beams, with energy in the range of 25-1000 keV, is due to collisions with protons and impurity ions (e.g., He$^{2+}$ and heavier impurities). As a result of such  excitation, at the particle density of 10$^{13}$-10$^{14}$ cm$^{-3}$, the line intensities of the Stark multiplets do not follow statistical expectations (i.e., the populations of fine-structure levels within the same principal quantum number $n$ are not proportional to their statistical weights). Hence, any realistic modeling of MSE spectra has to include the relevant collisional atomic data. In this paper we provide a general expression for the excitation cross sections in parabolic states within $n$=3 for an arbitrary orientation between the direction of the motion-induced electric field and the proton-atom collisional axis. The calculations make use of the density matrix obtained with the atomic orbital close coupling method and the method can be applied to other collisional systems (e.g., He$^{2+}$, Be$^{4+}$, C$^{6+}$, etc.). The resulting cross sections are given as simple fits that can be directly applied to spectral modeling. For illustration we note that the asymmetry detected in the first classical cathode ray experiments between the red- and blue-shifted spectral components can be quantitatively studied using the proposed approach.}


\begin{document}


\section{Introduction}
Beam-assisted spectroscopy represents a special class of diagnostics in plasma spectroscopy, as here, in contrast to passive emission spectroscopy, the heavy-particle collisions at the energies of a few atomic units  (1 a.u. $\approx$ 25 keV) play a dominant role \cite{Janev89}. Local measurements of ion temperature, concentration of impurity ions, plasma rotation, and electric field measurements including polarisation coherence, or control of the plasma current density profile, are the most representative examples of the use of injection of heating or diagnostic beams in fusion plasmas \cite{Isler77,Lev89, Del10, Ko16, ThormanPhD}. So, for instance,  charge-exchange recombination spectroscopy, which is based on the capture of bound electrons of the beam atoms by impurity ions, has been exploited for the last few decades on practically all former and present tokamaks and stellarators, including JET, ASDEX, W7-X, KSTAR, etc., \cite{vonHell05}. Beam-assisted spectroscopy is also expected to play a significant role in future fusion devices such as ITER \cite{vonHell18, Ralch19, Biel11}.

 The Motional (translational) Stark effect (MSE) diagnostic provides an excellent benchmark of atomic data for the simplest collisional systems, for example, excitation of H atoms by H$^+$ \cite{ Kuang96}. Both Stark effect measurements with beam atoms and Zeeman effect measurements for the cold atoms at the plasma edge can clearly detect the spectral line components using high resolution spectroscopy \cite{Car85, Mandl93}. For instance, for the applied external magnetic field of 1-5 T the energy separation due to the Zeeman effect is a few times larger than the fine-structure splitting of cold atoms in the plasma. In the rest frame of the beam atoms moving in the external magnetic field the bound electron experiences the static electric field $\vec{F}$ = $\vec{v}/c\times\vec{B}$ where $F$ and $B$ are the strengths of electric and magnetic field, respectively, $\vec{v}$ is the velocity of the beam atom, and $c$ is the speed of light. As can be easily seen for typical parameters of magnetic fusion plasmas, the energy separations induced by the Stark effect exceed the Zeeman splitting by a significant factor. Therefore, the representation of atomic structure for these specific conditions should primarily reflect the electric field effects, and in particular dictate usage of a parabolic basis. Accordingly, for MSE studies the parabolic quantum numbers offer a good description of the atomic structure and the emission of spectral lines \cite{Reimer16}.
 
Calculations of relative MSE line intensities in laboratory plasmas are based on either a statistical (static) or dynamical assumption \cite{BetheBook}. In the former case the populations of levels are considered to be proportional to their statistical weights. The line intensities are then derived using the following expression:
\begin{equation}
I_{a-b} \propto g_{a}A_{a-b},
\end{equation}
where $a={n_ak_am_a}$ and $b={n_bk_bm_b}$ are the sets of quantum numbers for the atomic $eigenstates$, $k=n_1-n_2$ is the electric quantum number, $n_1\ge0$ and $n_2\ge0$ are the parabolic quantum numbers and $m$ is the magnetic quantum number ($n=n_1 + n_2 + |m| + 1$); $A_{a-b}$ is the Einstein coefficient (radiative tranistion probability) for the transition between the states $a$ and $b$ \cite{Schroed26}. The dynamical intensities are calculated using the next formula:
\begin{equation}
I_{a-b} \propto g_aA_{a-b}/\sum_c{A_{a-c}}.
\end{equation}
It is generally accepted that Eq. (1) is valid for high density plasmas, whereas Eq. (2) is utilised to describe experiments at low densities where radiative relaxation occurs on timescales much faster than the collisional redistribution among the sub-levels. Though extensive atomic structure calculations were performed for both types of experimental conditions, the observed line intensities were not always adequately described using tabulated statistical or dynamical intensities. The disagreement among measured and calculated intensities accompanies the Stark effect observation starting from the dawn of quantum mechanics. The historical view of the problem was well characterized in Ref.~\cite{BookRyde}: \textit{"Upon the whole the problem of the hydrogen intensities showed a very confused picture"}.
The aim of this paper is to present a concise tabulated set of atomic data that has been found to be the most successful one, at least for fusion plasmas, by applying a density matrix formalism.

\section{Theoretical approach}

\begin{figure}
\centering
\includegraphics[scale=0.5,angle=0]{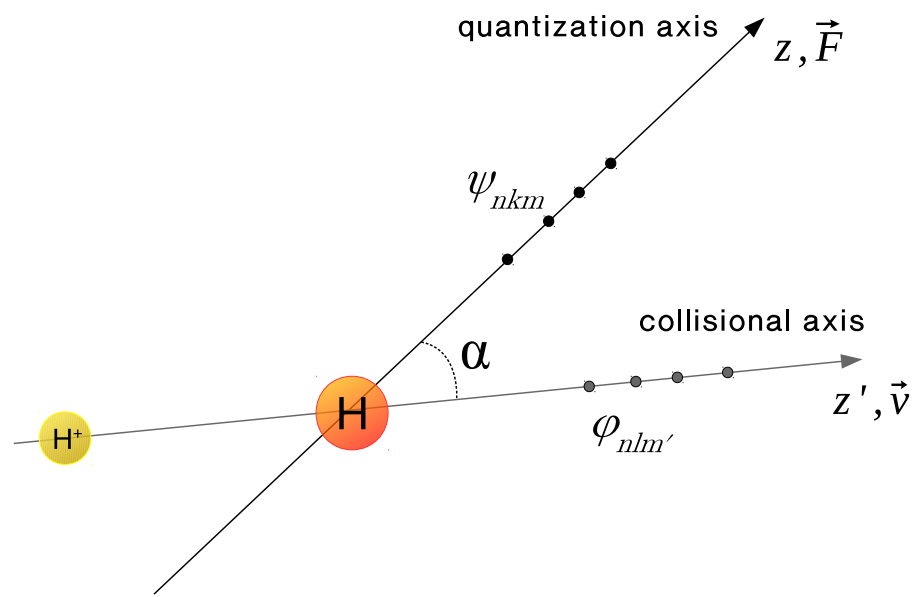}
\caption{Scheme of the transformation between spherical and parabolic eigenstates. The $z'$ axis defines the axis of symmetry of the collisional process in the absence of an electric field. The $z$ axis is the quantization axis defined by the direction of the electric field $\vec{F}$. For experiments in fusion plasmas, angle $\alpha$ is $\pi/2$ assuming the ions are cold (e.g., the ion temperature $T_i$ is much lower than the beam energy $E_b$).}
\label{fig-scheme}
\end{figure}

Measurements of hydrogen line intensities in the presence of electric fields are often related to anisotropic excitation. This is the case both for the classical cathode ray experiments and for the MSE measurements in fusion plasmas. Figure \ref{fig-scheme} illustrates the geometrical relationship between the collisional and electric field directions. In the absence of electric field the quantization axis is normally defined by the collision axis between incident ions and atoms. However, when an external electric field is applied to an atom, the spherical symmetry of an isolated atom is replaced by the axial symmetry defined by the direction of the electric field. 

For standard problems in atomic scattering, the quantization axis ($z'$) is normally selected along the projectile velocity. This choice results in well-defined spherical eigenstates of the atom $\phi_{nlm}$. However, for MSE conditions, the induced electric field provides a new quantization axis ($z$) which is to be used to define new parabolic eigenstates $\psi_{nkm}$ (see, e.g., Ref.~ \cite{March10} for details). Such modification requires calculation of collisional scattering amplitudes and cross sections for transitions between parabolic rather than spherical eigenstates. To utilize the standard techniques of atomic collision theory, one has to transform the parabolic basis wavefunctions quantized along $z$ to the spherical basis wavefunctions quantized along $z'$. This procedure actually requires two steps. The first one is a conventional quantum-mechanical rotation \cite{EdmondsBook} from axis $z$ to axis $z'$ by angle $\alpha$ and the second one is the transformation between the spherical states and parabolic states \cite{Park60,Herrick75} defined along the same quantization axis:
\begin{equation}
\psi_{nkm} = \sum_{l=|m|}^{n-1}{C^{lm}_{nk} \sum_{m'=-l}^{l}{d^{m'}_{lm}(\alpha) \varphi_{nlm'}} }
\label{transform}
\end{equation}
where $C^{lm}_{nk}$ is the Clebsch-Gordan coefficient and $d^{m'}_{lm}(\alpha)$ is the rotation matrix element. By applying formally the collision operator to the wavefunctions $\psi_{nkm}$ the excitation cross sections can be calculated using any applicable theoretical method, for example, atomic orbital close coupling (AOCC) \cite{Schultz15}, convergent close coupling (CCC) \cite{Abdur19}, or the Born or Glauber approximations \cite{Gu08, Tai70}.

General expressions for the  cross sections $\sigma_{nkm}$ for excitation from the ground state ($1s$) of hydrogen to the $n$=2 \cite{March11} and $n$=3 parabolic states $nkm$ are as follows:
\begin{align}
\sigma_{2\pm10} &= \frac{1}{2}\sigma_{2s0}+ \frac{1}{2}cos^{2}(\alpha)\sigma_{2p0}+ \frac{1}{2}sin^{2}(\alpha)\sigma_{2p1} \mp cos(\alpha)Re(\rho_{2s0}^{2p0}), \\
\sigma_{201} &= \frac{1}{2}sin^2(\alpha)\sigma_{2p0}+ \sigma_{2p1}\left(1-\frac{1}{2}sin^2(\alpha) \right ),\\
\sigma_{3\pm20} &= \frac{1}{3}\sigma_{3s0} + \frac{1}{4} \left ( 1 + cos(2\alpha)   \right ) \sigma_{3p0} + \frac{1}{4} \left (1-cos(2\alpha) \right )\sigma_{3p1}+  
\frac{1}{192} \left ( 11 +12cos(2\alpha)  + \right.    \nonumber \\ &\qquad 
\left. + 9cos(4\alpha) \right )\sigma_{3d0} + \frac{1}{16} \left ( 1 -cos (4\alpha)  \right)\sigma_{3d1} + \frac{1}{64}\left (3 -4cos (2\alpha ) + cos (4 \alpha ) \right) \sigma_{3d2} \mp \nonumber \\ &\qquad 
\mp \frac{\sqrt{6}}{3}cos(\alpha) Re(\rho_{3s0}^{3p0}) + \frac{\sqrt {2}}{12}\left(1 +3cos(2\alpha) \right) Re(\rho_{3s0}^{3d0}) \mp \nonumber \\ &\qquad
\mp \frac{\sqrt{3}}{24} \left( 3cos (3\alpha) + 5 cos (\alpha ) \right ) Re(\rho_{3p0}^{3d0}) \pm 
\frac{1}{4} \left ( cos (3\alpha) -  cos (\alpha ) \right)   Re(\rho_{3p1}^{3d1}), \\
\sigma_{3\pm11} &= \frac{1}{8}\left( 1-cos(2\alpha)  \right)\sigma_{3p0}  + \frac{1}{8}\left( 3 + cos(2\alpha)  \right)\sigma_{3p1} + \frac{3}{32}\left( 1 - cos(4\alpha)  \right)\sigma_{3d0}  + \nonumber \\ &\qquad
+\frac{1}{8}\left( 2 + cos(2\alpha)   + cos(4\alpha) \right)\sigma_{3d1} + 
\frac{1}{32}\left( 5 - 4 cos(2\alpha)  - cos(4\alpha) \right)\sigma_{3d2} \pm \nonumber \\&\qquad
\pm \frac{\sqrt{3}}{8}\left( cos(3\alpha) - cos(\alpha)  \right)Re(\rho_{3p0}^{3d0}) \mp \frac{1}{4}\left( 3cos(\alpha) + cos(3\alpha) \right)Re(\rho_{3p1}^{3d1}), \\
\sigma_{300} &= \frac{1}{3}\sigma_{3s0}+\frac{1}{48}\left(11 + 12 cos(2\alpha) + 9 cos(4\alpha)     \right)\sigma_{3d0} + \frac{1}{4} \left (1- cos(4\alpha) \right) \sigma_{3d1} + \nonumber \\&\qquad +\frac{1}{16}\left (3 -4cos (2 \alpha) + cos(4\alpha) \right )\sigma _{3d2}  - \frac{\sqrt{2}}{6} \left (1 + 3cos( 2\alpha ) \right) Re(\rho_{3s0}^{3d0}) ,  \\
\sigma_{302} &= \frac{3}{64} \left( 3 - 4cos(2\alpha) + cos(4\alpha)\right) \sigma_{3d0} + \frac{1}{16} \left( 5 - 4 cos(2\alpha) - cos(4\alpha) \right )\sigma_{3d1} + \nonumber \\ &\qquad
+ \frac{1}{64} \left( 35 + 28 cos(2\alpha) + cos (4\alpha) \right) \sigma_{3d2}. 
\label{eq-crosstrans}
\end{align}
Here $\sigma_{nlm}$ on the r.h.s (e.g., $\sigma_{3d0}$) is the excitation cross section from $n$=1 to the spherical state $nlm$ and $\rho^{nlm}_{n'l'm'}$ is the off-diagonal density matrix element. Note that $\sigma_{nkm} = \sigma_{nk-m}$ and $\sigma_{nlm}=\sigma_{nl-m}$ for $m\ne$0 for both parabolic and spherical states. Also, the sum of the excitation cross sections from $n$=1 to $n$=2 and $n$=3 equals that for the field-free case:
\begin{align}
\sigma_{n=2} &= \sigma_{s0} + \sigma_{p0} + 2\sigma_{p1} = \sigma_{210} + \sigma_{2-10}+ 2\sigma_{201}, \\
\sigma_{n=3} &= \sigma_{s0} + \sigma_{p0} + \sigma_{d0} + 2\left( \sigma_{p1} + \sigma_{d1} + \sigma_{d2} \right) = \sigma_{320} + \sigma_{3-20}+ \sigma_{300} + 2\left( \sigma_{3-11} + \sigma_{311} + \sigma_{302} \right) \;\; .
\label{eq-crosssum}
\end{align}

These expressions are valid for the linear Stark effect only, namely, when the energy splitting due to the induced electric field is much larger than the fine-structure splitting and much smaller than the energy separation between the levels belonging to different principal quantum numbers $n$. This condition is generally fulfilled only for levels with $n\lesssim5$ \cite{Lotte02, MarchSpringer}. 

Some physical properties can be inferred from equations (4-9) for excitation of parabolic states, which contain not only the $m$-resolved cross sections for the spherical states but also the off diagonal elements (real part) of the density matrix \cite{Blum79}. First, the cross sections exhibit a strong dependence on the angle between the direction of the electric field and the axis of symmetry of the collisional frame. This was remarkably demonstrated first in Refs.~\cite{Hickman83,Prunele85} through the study of excitation of highly excited circular Rydberg atoms ($k=n-1$) at thermal energies. It is the presence of the coherence terms in Eqs. (4-9) that explains the asymmetry between the red- and blue-shifted lines for Stark effect measurement (note that the influence of field ionisation is still low \cite{Damb79} under these conditions). Therefore, for instance, depending on the orientation between the cathode rays and the vector direction of the electric field (e.g., parallel, $\alpha=0$;  transverse, $\alpha=\pi/2$; or anti-parallel, $\alpha=\pi$) blue-shifted, red-shifted, or symmetric patterns were detected for Balmer-$\alpha$ components of the emission \cite{BookRyde}. In addition, it has to be mentioned that detailed  beam-foil experiments firmly established the value of employing the density matrix method in interpreting these measurements \cite{Ashburn89, Eck73}.

The first calculations of excitation cross sections in parabolic states relevant to fusion plasmas were performed using the Born approximation \cite{Gu08} and later the Glauber approximation \cite{March10}. It was explicitly shown how sensitive the ion-atom parabolic cross sections are to the orientation between the field direction and the ion-atom collision axis. However, at the collisional velocity of 1-2 a.u. none of perturbative two-state approximations could adequately describe the measured cross sections. This situation was improved by introducing the AOCC calculation for $n$=2 and $n$=3 states \cite{Schultz15}. Furthermore, the collisional radiative model in parabolic states was extended from $n$=5 to $n$=10 demonstrating the effect of field ionisation \cite{March11}. Recently new extensive CCC calculations of $m$-resolved cross sections and density matrix elements became available \cite{Abdur19}. Despite the somewhat stronger oscillations of the cross section as a function of collision energy, the magnetic-quantum-number resolved AOCC cross sections of Ref. \cite{Schultz15} and the new CCC results agree quite well\footnote{The AOCC excitation cross sections to 3d$_1$ and 3d$_2$ levels shown in Fig. 2 of Ref. \cite{Abdur19} are surprisingly lower than in the original publication \cite{March12}.}. 

Here we report accurate fits to the $m$-resolved cross sections (diagonal elements of the density matrix) and the coherence terms (off diagonal elements of the density matrix) of Ref. \cite{Schultz15} in the energy range of 20-2000 keV using the following formulas \cite{Jan93} : 
\begin{align}
\sigma(E) &= A_0 \left ( e^{-A_1/E}\frac{ln(1+BE)}{E} + A_2\frac{ e^{-A_3E}}{E^{A_4}}+A_5\frac{e^{-A_6/E}}{1+A_7E^{A_8}} \right ) ,\\
\sigma(E) &= A_0 \left ( e^{-A_1/E}\frac{1}{E} + A_2\frac{ e^{-A_3E}}{E^{A_4}}+A_5\frac{e^{-A_6/E}}{1+A_7E^{A_8}} \right ).
\label{eq-formula}
\end{align}

\noindent The energy E is in keV and cross sections are given in units of $\pi a_0^2 \approx$ 8.79$\cdot$10$^{-17}$ cm$^{2}$ where $a_0 \approx$ 0.529$\cdot$10$^{-8}$ cm is the Bohr radius. The derived coefficients are given in Table 1, and Figs. 2 and 3 show the original data and the fits for excitation to $n$=2 and the coherence terms for excitation to $n$=3, respectively.

\begin{table}
\begin{center}
\begin{tabular}{|p{0.6cm}| cccccccccc |}
 \hline
  & A$_0$ & A$_1$ & A$_2$ & A$_3$ & A$_4$ & A$_5$ & A$_6$ & A$_7$ & A$_8$ & B \\
 \hline
 2s$_0$ & 3.49+01 &3.05+01 &1.01-03 &5.33-01 &-2.78+00 &-9.33-03 &3.18+00 &1.07-02 &1.03+00 &0 
\\
 2p$_0$ & 5.51+01 &5.89+01 &9.31-05 &1.52-01 &-1.66+00 &1.45-05 &-2.69+00 &8.48-01 &-8.81+00 &0  \\
 2p$_1$ & 5.23+00 &7.37+01 &-6.02-34 &1.89-01 &-1.95+01 &2.79-02 &9.40+00 &1.86-02 &7.25-01 & 2.12+01
  \\
 -s$_0$p$_0$ & 4.96+00 &5.17+01 &0 &0 &0 &2.75-02 &7.84+00 &5.56-05 &2.29+00 & 0 \\
 \hline
 3s$_0$ & 3.79+00 &6.55+01 &1.19-04 &5.67-02 &-1.72+00 &-3.55-04 &6.25+01 &3.41+10 &-1.29+01 & 0 \\
 3p$_0$ & 9.52+00 &6.30+01 &4.03-05 &1.28-01 &-1.88+00 &-9.51-05 &4.04+03 &-7.13+12 &-1.85+01 & 0 \\
 3p$_1$ & 1.70-01 &6.37+01 &-6.92-01 &3.50-02 &1.80-01 &3.55-01 &2.38+00 &5.33-03 &9.40-01 & 1.14+01 \\
 3d$_0$ & 5.24-01 & 8.52+01 & 4.31-04 & 3.74-02 & -1.57+00 & 9.84-05 & -3.05+01 & 3.49+10 & -6.75+00 & 0\\
 3d$_1$ & 1.03-02 & 6.37+01 & -1.30+00 & 2.68-01 & -1.52+00 & 1.54+00 & 2.23+00 & 1.50-03 & 1.64+00 & 1.14+01\\
 3d$_2$ & 5.45-01 & 8.11+01 & 0 & 0 & 0 & 5.24-05 & -9.22+01 & 9.67+11 & -8.67+00 & 0 \\
 -s$_0$p$_0$ & 4.58+00 & 2.95+01 & 0 & 0 & 0 & -7.93-03 & -7.40-02 & 5.24-03 & 1.07+00 & 0\\
 s$_0$d$_0$ & 2.84e+00 & 2.73e+01 & 0 & 0 & 0 & -1.40e-02 & -7.71e-02 & 5.04e-03 & 1.20e+00 & 0 \\
 -p$_0$d$_0$ & 4.33+00 & 3.28+01 & 0 & 0 & 0 & -2.17-02 & -7.53-02 & 4.83-01 & 5.51-01 & 0\\
 -p$_1$d$_1$ & 5.25e-02 & 1.23e+02 & 0 & 0 & 0 & 3.03e+01 & 5.47e+01 & 1.34e-01 & 1.39e+00 & 0\\
 \hline
\end{tabular}
\caption { Table of fit coefficients for the excitation cross sections and the real part of coherence terms for excitation to $n$=2 and $n$=3. Expression (12) is applied to fit the AOCC data for all elements except the 2p$_1$, 3p$_1$, and 3d$_1$ cross sections in which case the expression (13) must be used. We note that the real part of all coherence terms is negative except for $s_0d_0$ excitation.}
\end{center}
\end{table}

\begin{figure}
\centering
\includegraphics[scale=0.6,angle=270]{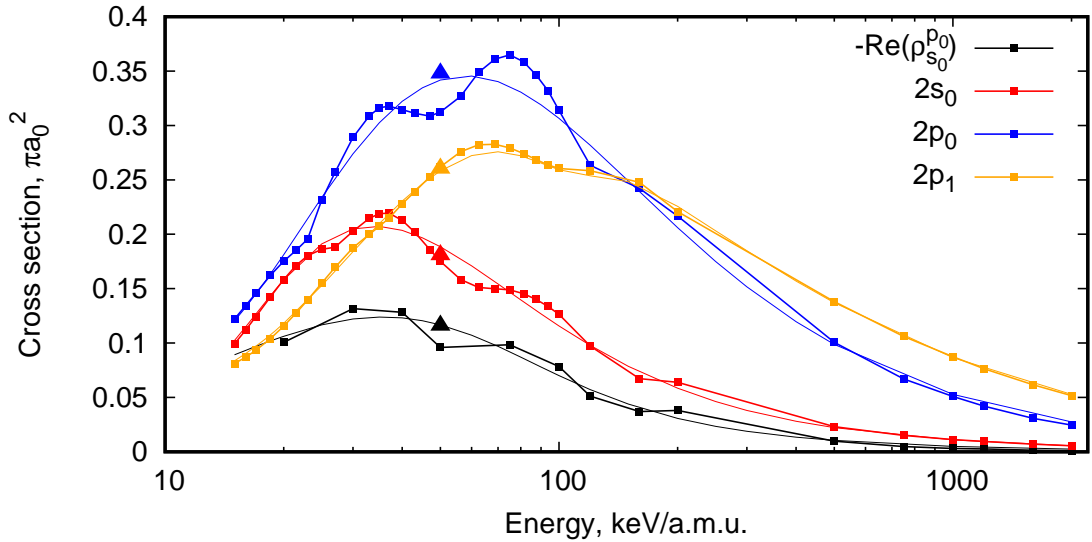}
\caption{Excitation cross sections for $n$=2 states and real part of the coherence term \cite{Schultz15} (thick lines with squares). The results of the fit are shown as thin lines using the same colours. The new CCC data at 50 keV are shown as triangles using the same colours (Table 1 of Ref.~\cite{Abdur19}).}
\label{fig-n=2}
\end{figure}

\begin{figure}
\centering
\includegraphics[scale=0.6,angle=270]{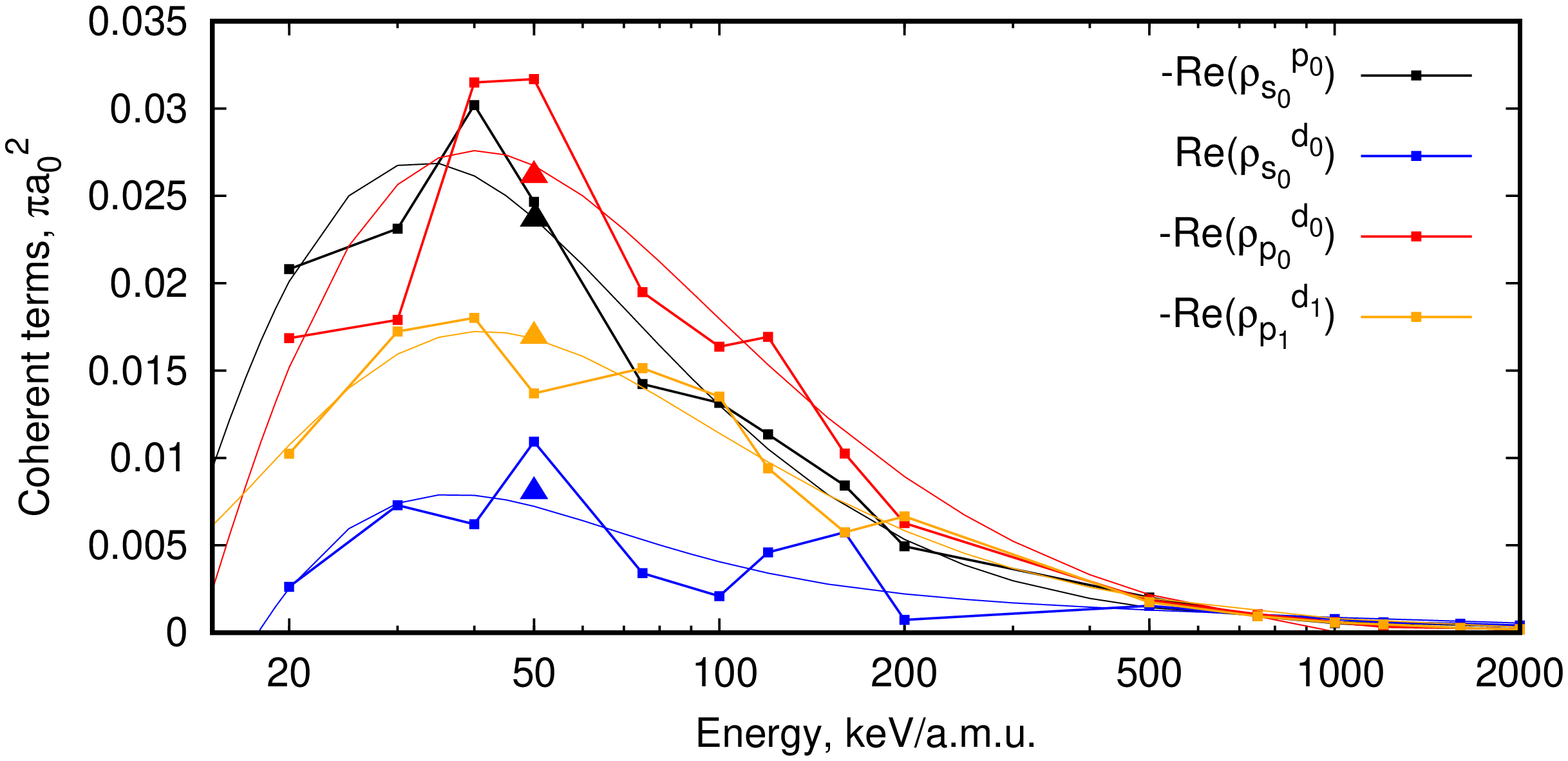}
\caption{Real parts of the excitation coherence terms to $n$=3 (thick lines with squares). The results of the fit are shown as thin lines with the same colours. The new CCC data at 50 keV are shown as triangles using the same colours (Table 1 of Ref.~\cite{Abdur19}).}
\label{fig-n=3}
\end{figure}

Using the data of Table 1 and Eqs. (4-9) one may calculate the excitation rate coefficients to parabolic states even at conditions when the ion temperature is comparable to the beam energy, and also analyse the Stark effect spectrum at an arbitrary angle of electric field direction.

\section{Lines ratio of Stark components in the low density limit}

We analyse in detail here the influence of the coherence elements on the line ratios of Stark components. For this purpose the collisional-radiative (CR) model NOMAD \cite{Ralch01} was adopted. The analysis has been performed in the low density limit, i.e., neglecting both the collisional redistribution between the excited states and the collisional ionisation. We also neglect the effect of radiative cascades above $n$=3. This approximation was used to better highlight the effect of direct-excitation coherent terms on line intensities. It is expected to be generally valid for densities below about (1-5) $\times$ 10$^{12}$ cm$^{-3}$. For denser plasma simulations, a CR model has to include the above mentioned processes. 

\begin{figure}[H]
\centering
\includegraphics[scale=0.7,angle=270]{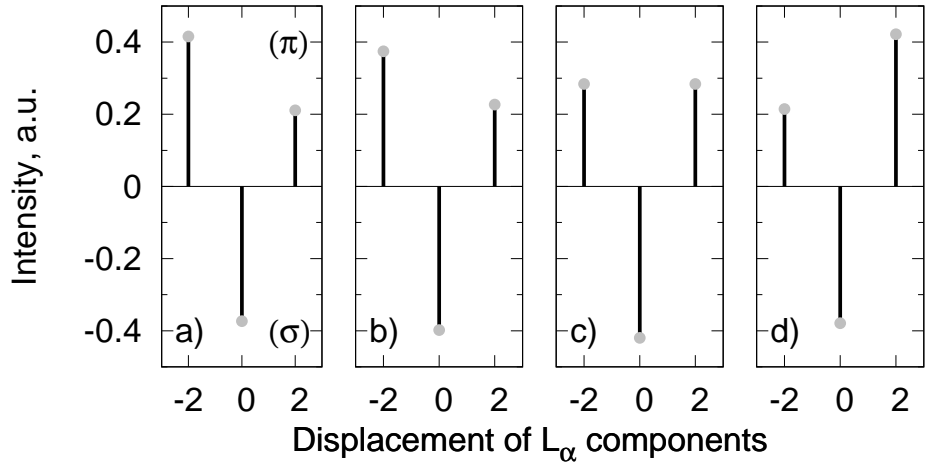}
\caption{The Stark effect for the Lyman-$\alpha$ line excited by proton impact at collision energy of 50 keV/u and by electrons with  temperature of 3 keV. The results of the calculation are shown for angles $\alpha=0$ (a), $\alpha=\pi/4$ (b), $\alpha=\pi/2$ (c) and $\alpha=\pi$ (d) between the axis of collision and the direction of electric field taking into account the cascades from $n$=3. For the statistical intensities the ratio between components $\pi_2^{\pm}:\sigma_0$ = 1:2, and for the dynamical intensities all three components are equal. The $\sigma$ component is shown with negative sign for better visibility. The displacement is given in units of $\frac{3\hbar F}{8\pi^2 m_e e c}$, where $\hbar$ is a Plank constant, $F$ is the strength of electric field, $m_e$ is the mass of electron, $e$ is the electron charge, and $c$ is the speed of light.}
\label{fig-La}
\end{figure}

\begin{figure}[h]
\centering
\includegraphics[scale=0.7,angle=270]{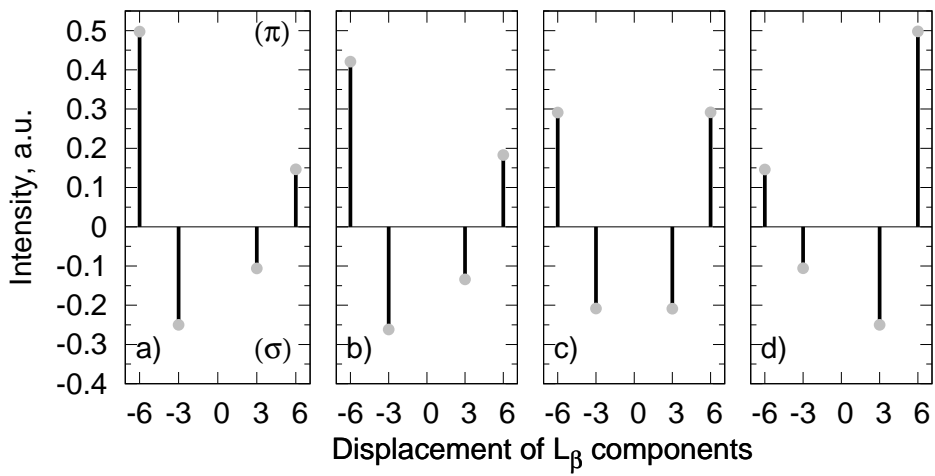}
\caption{The Stark effect for the Lyman-$\beta$ line excited by proton impact at a collision energy of 50 keV/u and by electrons with  temperature of 3 keV.  For the statistical intensities the ratio between components $\sigma_3^{\pm}:\pi_6^{\pm}$ = 25:25, and for the dynamical intensities the ratio is 23:27. Other notations are the same as in Figure \ref{fig-La}.}
\label{fig-Lb}
\end{figure}

\begin{figure}
\centering
\includegraphics[scale=0.6,angle=270]{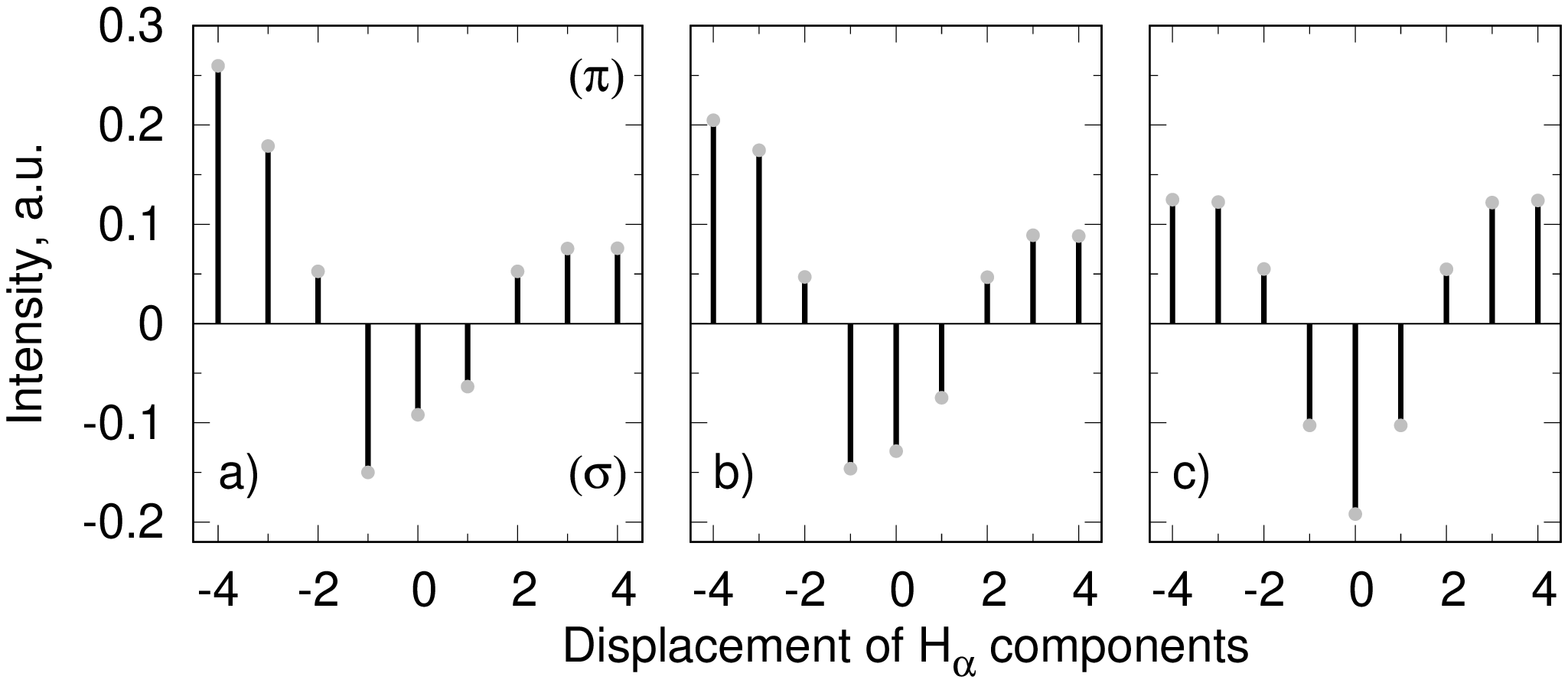}
\caption{The Stark effect for the Balmer-$\alpha$ line excited by proton impact at a collision energy of 50 keV/u and by electrons with temperature of 3 keV. For the statistical intensities the ratio between components $\pi_4^{\pm}:\pi_3^{\pm}:\pi_2^{\pm}:\sigma_1^{\pm}:\sigma_0$ = 17.8:24.4:7.7:20.5:58.2, and for the dynamical intensities the ratio equals to 15.6:18.2:16.2:12.5:74.6. The weak transitions $\sigma_5^{\pm}$, $\sigma_6^{\pm}$ and $\pi_8^{\pm}$ are omitted from the graph. Other notations are the same as in Figure \ref{fig-La}. }
\label{fig-Ba}
\end{figure}

The calculations were carried out for the neutral beam energy of 50 keV/u and electron temperature of 3 keV. The density of ions and electrons were equal in the calculations. Here we report the line intensities of the Lyman-$\alpha,\beta$ and Balmer-$\alpha$ lines for different values of $\alpha$ (Figs. \ref{fig-La}-\ref{fig-Ba}). The angle between the line-of-sight and the direction of the electric field is $\theta=\pi/2$. In this case both $\sigma$ ($\Delta$ m=$\pm$1) and $\pi$ ($\Delta$ m = 0) transitions are observable. In the statistical limit, $\sum_{\sigma}{I_\sigma} = \sum_{\pi}{I_\pi} = 1/2$. Figure \ref{fig-La} shows the results of the calculation for the Lyman-$\alpha$ line at four different values of $\alpha$. The most striking effect is a strong asymmetry for the blue- and red-shifted components at $\alpha \ne \pi/2$. Indeed, depending on the angle between the field and the projectile velocity (parallel or anti-parallel) either the blue- or the red- shifted $\pi$ component dominate the spectrum. The blue-shifted component is a factor of two higher compared to the red-shifted one in Fig. \ref{fig-La}.a. In Figure \ref{fig-La}.b, at the angle $\alpha=\pi/4$, the ratio is reduced only marginally. In Fig. \ref{fig-La}.c, with $\alpha=\pi/2$ that corresponds to the MSE conditions in fusion plasmas, the intensities of both components are equal. Finally, in Fig. \ref{fig-La}.d the red-shifted $\pi$ component is stronger than the red-shifted one. As mentioned above, this asymmetry between the components $\pi_2^{\pm}$ is due to the presence of the coherence term $Re(\rho_{2s_0}^{2p_0})$ in the expression for the cross sections. 

Experimental evidence of such asymmetry was obtained from cathode ray experiments under much more complex experimental conditions. In this case, the line emission is affected by electron-impact excitation, dissociation of molecules, and charge exchange between protons and atoms. Nevertheless, the physical picture of the asymmetry is attributed to screening of the trajectories of bound electrons from incident protons or electrons by nucleons at rest as was already pointed out by Bohr \cite{Bohr15}: the charge distribution of the majority of parabolic states is asymmetrical relative to the plane with $z$ = 0 (see, for instance, Figure 8 of Ref.~\cite{BetheBook}). In this case the states with $k$=$n_1$-$n_2$>0 cannot be as efficiently excited as the states with $k<$0 for a proton travelling along the $z$ axis from -$\infty$ to +$\infty$. Because the direction of the electric field points against the direction of cathode rays (e.g., in the opposite direction to the protons) it leads to the dominance of the red-shifted emission (Figure \ref{fig-La}.d). In contrast to the strong variation of the $\pi_2^\pm$ component the variation of the $\sigma_0$ component is negligibly small. Its value only varies from 0.38 to 0.42. For all excitation angles $\sum_{\sigma}{I} / \sum_{\pi}{I} < 1$. We note that observed asymmetry in Lyman-$\alpha$ spectrum cannot be reproduced using a simple two-state approximation. Calculations using the Born or Glauber approximations give $Re\left( \rho_{s_0}^{p_0}\right )=0$, so that the symmetrical picture is expected in a MSE observation \cite{Schoell86, March10}. In the latter case the ratio between the the $\pi_2^\pm$ and $\sigma_0$ lines is closer to the dynamical limit.\\

Similar behaviour is found for the Lyman-$\beta$ line (Fig. \ref{fig-Lb}). Here, as in the case of Lyman-$\alpha$ emission, the variation of intensity between the blue- and the red-shifted $\pi_6$ and $\sigma_3$ components is found. The ratio between the $\sigma$ and $\pi$ components for MSE conditions is quite close to the dynamical limit. The ratio is found to be 21:29 against 23:27. As for excitation by proton-impact of the Lyman-$\alpha$ line, $\sum_{\sigma}{I} / \sum_{\pi}{I} < 1$ for all angles with the electric field direction.

For the Balmer-$\alpha$ line (Fig. \ref{fig-Ba}), however, the $\pi_2^\pm$ line remains rather insensitive to variation of the angle between the incident protons and the direction of the electric field. This component originates from the state (300) that contains only the even term $Re(\rho_{3s0}^{3d0})$ so that this component remains symmetrical for any angle. Also for this line $\sum_{\sigma}{I_\sigma} / \sum_{\pi}{I_\pi} < 1$. In the low density limit for MSE conditions the ratios are $I_{\sigma_1}/I_{\sigma_0}$ = 0.53, $I_{\pi_4}/I_{\pi_3}$= 1, and $I_{\pi_2}/I_{\pi_3}$ = 0.45. It should be noted that the principal mechanism of excitation is via proton collisions while electrons contribute at the level of only 20-30\%. Nevertheless, not all the states within $n$=3 can be excited by electrons at temperatures of 100 eV or more. For instance, the ratio $\left( I_{\sigma_1}/I_{\sigma_0} \right)$>0.35 is due to preferential excitation of the states (3$\pm$11) by electrons. The states (302) and (300) cannot be excited by electrons at all. For this reason the electrons should not be neglected completely in the analysis of the data as they lead in general to a different distribution than from unidirectional proton impact.

\section{Conclusion}

In  this work we provided a set of proton-hydrogen excitation cross sections for calculation of the line intensities for radiative transitions between $n \leq 3$ parabolic states in the low-density limit. These data are required for MSE diagnostics under typical conditions of fusion plasmas. We expect that the lines ratio in the low density limit should be valid up to the electron densities of (1-5) $\times$ 10$^{12}$ cm$^{-3}$ depending on the beam energy. For denser plasmas, however, one has to develop a full collisional-radiative model with account of collisional and radiative transitions between excited states. Unfortunately, it is not easy to represent the cross sections for transitions with  $\Delta$n=0 in closed form as they also depend on the magnetic field. 

The exact expressions for the cross sections in parabolic states for $n$=3 excitation are given for the first time for an arbitrary angle between projectile velocity and direction of electric field. The expressions include, in particular, coherence terms of the density matrix. Using this approach one can efficiently describe the excitation for typical MSE conditions. In addition, this method can be used to model the asymmetry of spectral-line emission observed, for instance, in cathode ray experiments where ionisation by the electric field can still be neglected. We also note that the asymmetry observed in the line shape of MSE, e.g. the obvious difference between $\pi^{\pm}$ components is connected with the geometry of observations, for instance the beam width on the order of tens of centimeters results in a stronger attenuation of edge beamlets. At the same time the different values of magnetic field at both positions (assuming the observation port at the outer wall of fusion device) results in a shape asymmetry of all the spectral components. We plan to analyze the recently measured KSTAR MSE spectra \cite{Ko16} in the nearest future as well to provide the data for excitations by He$^{2+}$, Be$^{4+}$ and C$^{6+}$.

In general, the modeling of Stark effect stimulated by electron excitation in gas discharges also requires density matrix calculations for dissociative recombination, electron-impact excitation, and charge exchange. Moreover, to the best of our knowledge, such calculations are not available or were not required until now. We hope that the present work will motivate such calculations in the future. 


\funding{This research was funded by the Program-oriented Funding (PoF) of the Helmholtz-Gemeinschaft Deutscher Forschungszentren (HGF).}

\acknowledgments{This work was carried out within the framework of the EUROfusion Consortium and has received funding from the Euratom research and training program 2014--2018 and 2019--2020 under Grant agreement No. 633053. The views and opinions expressed herein do not necessarily reflect those of the European Commission.}

\conflictsofinterest{The authors declare no conflict of interest} 
\reftitle{References}





\end{document}